\documentclass[apj,twocolumn,twocolappendix,numberedappendix]{openjournal}
\usepackage{newtxtext,newtxmath,enumitem,multirow,ctable,amsmath,sidecap}
\usepackage[T1]{fontenc}
\usepackage[breaklinks,colorlinks,citecolor=blue,urlcolor=blue]{hyperref}

\newcommand{\orcidauthor}[3]{\author{\href{http://orcid.org/#1}{#2$^{#3}$}}}

\shorttitle{UV-compactness of high-$z$ galaxies}
\shortauthors{Langeroodi \& Hjorth}

\begin{document}

\title{\vspace{-0.8cm}Ultraviolet Compactness of High-Redshift Galaxies as a Tracer of Early-Stage Gas Infall, Bursty Star Formation, and Offset from the Fundamental Metallicity Relation\vspace{-1.5cm}}

\orcidauthor{0000-0001-5710-8395}{Danial Langeroodi}{1,*}
\orcidauthor{0000-0002-4571-2306}{Jens Hjorth}{1}
\affiliation{$^{1}$DARK, Niels Bohr Institute, University of Copenhagen, Jagtvej 155A, 2200 Copenhagen, Denmark}

\thanks{$^*$E-mail: \href{mailto:danial.langeroodi@nbi.ku.dk}{danial.langeroodi@nbi.ku.dk}}

\begin{abstract}
The empirical anti-correlation between gas-phase metallicity and star formation rate (SFR) at a fixed stellar mass, known as the fundamental metallicity relation (FMR), is commonly interpreted as an equilibrium state in the interplay between gas infall, enrichment, and gas removal. JWST/NIRSpec spectroscopy has shown a $z>3$ deviation from the local-universe FMR calibrations, suggesting that these galaxies are potentially caught out of equilibrium. To investigate this, we inferred the stellar population, nebular, and morphological properties of 427 galaxies at $3<z<10$ using uniformly reduced NIRSpec prism spectroscopy and NIRCam photometry. We consider morphology as a possible indicator of chemical enrichment stage. We find a mass-size relation already in place at $4<z<10$, with a normalization anti-correlated with redshift. The size-redshift anti-correlation depends on stellar mass: while the size of $M_{\star}<10^8M_{\odot}$ galaxies strongly declines with redshift, $M_{\star}>10^9M_{\odot}$ galaxies exhibit negligible redshift evolution. We also confirm the redshift evolution of the FMR: $z>3$ galaxies appear metal-deficient compared to expectations for their stellar mass and SFR according to the local-universe FMR. This offset grows with redshift. Metal deficiency is correlated with compactness: galaxies most offset from the average mass-size relation are also the most metal-poor for their stellar mass and SFR. We interpret this as a product of bursty star formation: compact galaxies exhibit elevated SFR surface densities, indicating that they are observed during burst episodes triggered by gas infall. While accretion of metal-poor gas has reduced their gas-phase metallicity by diluting the interstellar medium, they are observed prior to chemical yield release by newly formed massive stars. Simply, they are chemically out of equilibrium compared to the equilibrium state known as the FMR.
\vspace{8pt}
\end{abstract}


\section{Introduction} \label{sec: intro}

The empirical correlation between the gas-phase metallicity and stellar mass of galaxies (i.e., the mass-metallicity relation) places strong constraints on theories of galaxy formation and evolution. The shape and normalization of this relation are thought to be determined by the interplay between gas infall, star formation, chemical enrichment, and feedback-driven outflows. This makes the mass-metallicity relation and its redshift evolution some of the most fundamental observables often attempted to reproduce by semi-analytic models and numerical simulations \citep{2015MNRAS.446..521S, 2016MNRAS.461.1760H, 2016MNRAS.459.2632L, ma+2016, 2017MNRAS.472.3354D, 2019MNRAS.484.5587T, 2020MNRAS.494.1988L, 2023MNRAS.518.3557U, 2023arXiv230503753H}.

Ground-based spectrographs have been used to probe the mass-metallicity relation out to $z \approx 3.5$, revealing a significant redshift evolution of its normalization \citep{sanders+2015, sanders+2021}. At a fixed stellar mass, $z \approx 3.5$ galaxies are on average $2-3$ times less metal enriched than those in the local universe. Probing the mass-metallicity relation at higher redshifts with ground-based instruments becomes challenging since the rest-optical metallicity diagnostic lines get redshifted out of the atmospheric transmission window. At $z > 3$, ALMA-accessible far infrared [\ion{O}{3}]$52\mu$m and [\ion{O}{3}]$88\mu$m lines can be used as alternatives to rest-optical metallicity diagnostic lines \citep{jones+2020}. However, compiling large samples of such observations has proven challenging due to the required integration time and the lack of secure spectroscopically-confirmed targets. This is likely to change however, since ALMA follow-ups of large samples of NIRSpec-confirmed $z > 3$ galaxies are underway.

JWST NIRSpec \citep{2022A&A...661A..80J} provides rest-optical spectroscopy for intermediate- to high-redshift galaxies, from $z \approx 3$ out to $z \approx 10$. During the first JWST cycle, this has already enabled new constraints on the mass-metallicity relation at $z > 3$ \citep[e.g.][]{langeroodi2023, 2022arXiv221202890H, 2022A&A...665L...4S, 2022ApJ...940L..23A, 2022ApJ...939L...3T, 2023ApJ...945...35T, 2023ApJ...942L..14R, 2023MNRAS.518..425C, 2023ApJ...950...67M, Nakajima+2023, curti+2023}. \cite{langeroodi2023} presented the first constraints on the mass-metallicity relation at $z \approx 8$, inferring significant evolution in its normalization with respect to the local universe and a mild evolution with respect to the $z \approx 2-3$ constraints. Other studies exploiting larger samples of $z > 3$ galaxies compiled from the CEERS \citep{ceers1, 2023arXiv230107072T, 2023ApJ...949L..25F} and JADES \citep{JADES1} NIRSpec observations have probed the mass-metallicity relation at $z = 3-8$ \citep{Nakajima+2023, curti+2023}. A consistent picture is starting to emerge: there is a mild redshift evolution in the normalization of the mass-metallicity relation at $z > 3$, and the slope of the mass-metallicity relation seems to be flattening with increasing redshift. 

The redshift evolution of the mass-metallicity relation out to $z \approx 3$ is consistently explained by the fundamental metallicity relation \citep[FMR; see][]{AM13, curti+2020}. The FMR is an empirical anti-correlation between the gas-phase metallicity and star-formation rate (SFR) at a fixed stellar mass. Since at fixed stellar mass, SFR increases with redshift \citep[i.e., the redshift evolution of the star-forming main sequence; see e.g.,][]{2023MNRAS.519.1526P}, the FMR predicts a decline in the normalization of the mass-metallicity relation with increasing redshift. This has been fully capable of capturing the redshift evolution of the mass-metallicity relation out to $z \approx 3$. 

The FMR remains ``fundamental'' out to cosmic noon: there is no evidence for a redshift-evolving FMR out to $z \approx 3$ \citep[e.g. see][]{sanders+2021}. However, recent NIRSpec studies have shown that the FMR exhibits some evolution at $z > 4$ \citep{2022arXiv221202890H, Nakajima+2023, curti+2023}. The inferred gas-phase metallicities of $z > 4$ galaxies seem to be offset from what is expected for their stellar mass and SFR, based on the local universe calibration of FMR. This offset seems to increase with increasing redshift. In simple terms, high-redshift galaxies appear more metal-poor than expected if the local universe calibration of the FMR were to hold. This indicates that the FMR is not fully capable of capturing the redshift evolution of the mass-metallicity relation at redshifts beyond $z \approx 4$.

The main mechanisms driving the redshift evolution of the FMR are not yet understood. However, there are speculations that the offset from the local universe calibration of FMR is correlated with the compactness of galaxies. The main piece of evidence was presented in \cite{tacchella+2023}, finding that the most metal-poor $z > 7$ galaxy detected in the Early Release Observations \citep{ERO} data is extremely compact ($R_{\rm e} < 200 {\rm pc}$). This galaxy has a high SFR surface density, potentially indicating that it is undergoing rapid/early-stage accretion, consistent with its inferred steeply-rising star-formation history (SFH) and relatively young stellar population \citep{langeroodi2023}. In this context, the offset from the FMR can be understood as the earliest stage of galaxy formation when the star formation is bursty \citep{2023arXiv231016895B, 2020MNRAS.498..430I, 2023MNRAS.525.2241H, 2024ApJ...961...53I, 2024MNRAS.52711372A, 2024ApJ...964..150D, 2024MNRAS.529.4728D, 2024arXiv240413045L} and equilibrium between gas infall, star formation, interstellar medium (ISM) enrichment, and feedback-driven outflow is not yet achieved. This equilibrium state is likely what we observe as the FMR for more mature galaxies at lower redshifts. 

Here, we investigate the correlation between the offset from the FMR and compactness by compiling a large sample of $z > 3$ galaxies with available rest-UV to -optical NIRSpec spectroscopy and NIRCam photometry. The former enables empirical method metallicity measurements and the latter enables accurate rest-ultraviolet (UV) size measurements. We compiled this sample by uniformly reducing the entire archive of publicly available JWST NIRSpec MSA prism data, resulting in 334 emission-line galaxies at $z > 3$. We also leverage the first public data release of the JADES program \citep{JADES1, JADES2, JADES3, JADES4}, providing an additional 93 galaxies at $z > 3$. A detailed overview of the public data used in this work as well as our reduction routines are presented in Section \ref{sec: data}. 

We describe our methods for measuring the stellar masses, sizes, metallicities, and SFRs of the galaxies in our sample in Sections \ref{sec: phot analysis} and \ref{sec: spec analysis}. In Section \ref{sec: size evol}, we present our inference of the mass-size relationship and its redshift evolution and introduce our measure of compactness. In Section \ref{sec: FMR}, we present our inference of the redshift evolution of the FMR. We also investigate the correlation between compactness and the offset from the local universe calibration of FMR. We discuss our findings and conclude in Sections \ref{sec: discussion} and \ref{sec: conclusion}. 

Throughout this work we adopt a standard $\Lambda$CDM cosmology with $H_{0} = 70$ km s$^{-1}$ Mpc$^{-1}$, $\Omega_\textnormal{m}$ = 0.3, and $\Omega_{\Lambda}$ = 0.7. Furthermore, we adopt a \cite{chabrier+2003} stellar initial mass function (IMF). Magnitudes are reported in the AB system \citep{1983ApJ...266..713O}.

\section{Data} \label{sec: data}

We reduced the entire publicly available cycle 1 NIRSpec prism multi-object spectroscopy \citep[MOS;][]{2022A&A...661A..80J, 2022A&A...661A..81F, 2023PASP..135c8001B}. We constructed a sample of 427 galaxies at $z_{\rm spec} > 3$ with secure spectroscopic redshifts based on high-significance (S/N $>3$) detections of the rest-optical metallicity diagnostic lines [\ion{O}{2}]$\lambda \lambda 3727, 3729$\AA\ doublet, H$\beta$, and [\ion{O}{3}]$\lambda \lambda 4959, 5007$\AA\ doublet. We limited our selection to galaxies with NIRCam or HST photometry coverage. Our sample consists of 334 galaxies from public data. In this Section we provide an overview of the corresponding NIRSpec spectroscopy and JWST/HST imaging, as well as the routines deployed to reduce them. A more detailed overview of our reduction pipelines is presented in \cite{2024arXiv240413045L}. The remaining 93 galaxies were selected from the first public data releases of the JADES program \citep{JADES1, JADES2, JADES3, JADES4}. 

\subsection{Spectroscopy} \label{sec: data-spec}

The NIRSpec MOS prism data were obtained through programs 1345 (CEERS), 1433, 2750, 2756, and 2767. We retrieved the raw data products (\texttt{uncal.fits}) and micro-shutter assembly (MSA) files (\texttt{msa.fits}) from the Barbara A. Mikulski Archive for Space Telescopes (MAST)\footnote{\url{https://mast.stsci.edu}}. We performed the detector-level corrections and converted the raw data into count-rate images (ramps-to-slopes) using the \texttt{Detector1Pipeline} routine of the official STScI JWST pipeline \citep{2022A&A...661A..81F}. We used the 1.10.0 version of the pipeline and the \texttt{jwst\_1077.pmap} Calibration Reference Data System (CRDS) context file. 

The 2nd and 3rd reduction levels were carried out using the \texttt{msaexp} package \citep{msaexp}\footnote{\url{https://github.com/gbrammer/msaexp}}, a python package employing the official STScI pipeline with additional custom routines for further corrections. In particular, these routines pre-process the rate images to correct for the residual $1/f$ noise; detect and remove the ``snowball" artifacts caused by large cosmic ray impacts; and remove bias, exposure by exposure. The WCS registration, flat-fielding, slit path-loss corrections, and flux calibration are performed using the \texttt{AssignWcs}, \texttt{Extract2dStep}, \texttt{FlatFieldStep}, \texttt{PathLossStep}, and \texttt{PhotomStep} routines from the \texttt{Spec2Pipeline} module of the STScI pipeline, respectively. 

The background is subtracted locally using a three-shutter node pattern, before drizzling the background-subtracted images onto a common grid. \texttt{Msaexp} extracts the optimal 1D spectra following the methods proposed in \cite{1986PASP...98..609H}: the 2D spectrum is integrated along the cross-dispersion axis, and then the resulting signal along the spatial axis is fitted with a Gaussian profile to measure the spatial offset and aperture of the inverse-variance weighted kernel that extracts the optimal 1D spectrum. 

For each galaxy, we performed further flux calibration by re-scaling its 1D spectrum to its measured multi-band NIRCam photometry (see Section \ref{sec: data-phot} for the details of NIRCam reduction and photometry measurements). For each galaxy, we measured the ratio of NIRCam to NIRSpec flux in each available filter, using the \texttt{sedpy} package \citep{sedpy} to project the 1D NIRSpec spectra onto NIRCam filters. The median ratio is defined as the flux calibration factor, used to re-scale the 1D NIRSpec spectra. 

We fitted the 1D spectra with \texttt{EAZY} \citep{eazy} to measure the spectroscopic redshifts. The fitted spectra were visually inspected to confirm the measured redshifts and to select the galaxies which satisfy the selection criteria mentioned above. 

\subsection{Photometry} \label{sec: data-phot}

The NIRCam and HST photometry of the fields targeted by the NIRSpec MOS observations considered in this work were acquired from the Grizli Image Release (v6.0) repository\footnote{The images are available at \url{https://grizli.readthedocs.io/en/latest/grizli/image-release-v6.html}}. This repository provides consistent reductions of the publicly available cycle 1 NIRCam imaging as well as the ancillary HST data (G. Brammer, in prep.). This includes mosaics of the EGS field (program 1345), as well as the fields toward lensing clusters Abell 2744 (programs 1324, 2561, and 2756), MACSJ0647 (program 1433), and RXJ2129 (program 2767). The NIRCam photometry was calibrated using the \texttt{jwst\_0995.pmap} CRDS context file. The stray-light features (i.e., ``wisps") were subtracted and the striping was removed using the \texttt{Grizli} software \citep{grizli}. All the NIRCam and HST images were aligned to a common reference image and drizzled to the same 40 mas pixel grid. 

Point spread function (PSF)-matching is required for measuring accurate source colors that are unaffected by the filter/wavelength-dependent size of the PSF. For this purpose, we PSF-matched the imaging of every filter that has a PSF FWHM below that of F444W, to the PSF in this filter. For each field, the empirical PSF in each filter was constructed using the \texttt{PSFex} package \citep{psfex}. Potential stars were identified based on their location on the half-light-radius vs source magnitude plane, produced using the \texttt{Source Extractor} software \citep[\texttt{SE};][]{sex}. The unsaturated stars were selected and pruned against significant contaminants and high fractions of bad pixels. These stars were re-centered, stacked, and then normalized to construct the empirical PSFs. PSF-matching kernels from the bluer filters to F444W were produced following the JWST post-pipeline Data Analysis Tools Ecosystem routines\footnote{The routines are available at \url{https://spacetelescope.github.io/jdat_notebooks/notebooks/NIRCam_PSF-matched_photometry/NIRCam_PSF_matched_multiband_photometry.html}}, and the mosaics in these bands were convolved to match the PSF in the F444W filter. 

The PSF-matching procedure was not applied to the filters where PSF FWHM is larger than that of the F444W. This includes the NIRCam F480M filter as well as the HST WFC3\_IR F105W, F110W, F125W, F140W, and F160W filters. We corrected the fluxes measured in HST WFC3\_IR filters by multiplying them by a factor of 1.25 as suggested in \cite{2022ApJ...928...52F} \citep[see also][]{ceers1}, derived through source-injection simulations \citep[note that this correction is specific to the HST WFC3\_IR filters; below we apply another 1.08 correction to the flux measured in all the filters, which makes the overall correction consistent with the 1.35 factor reported in][]{ceers1}.

Multi-band photometry of the spectroscopically-selected galaxies in our final sample were measured in $0.3\arcsec$ circular apertures (diameter) by running \texttt{SE} in dual-image mode. The detection and deblending criteria were set to \texttt{DETECT\_MINAREA} = 5, \texttt{DETECT\_THRESH} = 3.0, \texttt{DEBLEND\_NTHRESH} = 32, \texttt{DEBLEND\_MINCOUNT} = 0.005. The detection image in each field was constructed from the variance-weighted combination of the F277W, F356W, and F444W mosaics. In the few cases where the galaxy was not covered in NIRCam imaging, the F160W image was used for detection. For each galaxy, $1\arcsec \times 1\arcsec$ cutouts of the detection image and segmentation map were visually inspected to ensure accurate detection and sufficient deblending. 

We used the \texttt{MAG\_APER} measurements on the PSF-matched images as the measured magnitudes and the \texttt{MAGERR\_APER} measurements on the original images (i.e., before PSF-matching) as their uncertainties. To perform aperture correction, we derived the correction factor for each galaxy as the offset between \texttt{MAG\_AUTO} and \texttt{MAG\_APER} magnitudes in the F444W image. We scaled the measured \texttt{MAG\_APER} values in all the filters by this factor to account for the flux that is not captured by the chosen aperture. We also corrected for the $\sim 0.03$ mag systematic offset between the \texttt{MAG\_AUTO} values and the true total magnitudes, resulting from the wings of the PSF not being captured by the \texttt{MAG\_AUTO} measurements \citep[see, e.g.,][]{ceers1}. Moreover, to account for further uncertainties (e.g., zero points), we forced a noise floor of $10\%$ on the measured magnitudes in each photometry band \citep[see, e.g.,][]{ceers1, harikane}.

We corrected for Galactic extinction using the E(B-V) reddening values from \cite{2011ApJ...737..103S}, assuming a \cite{CCM89} attenuation curve. This corresponds to E(B$-$V) = 0.0089, 0.0112, 0.0966, and 0.0349 mag, respectively for the EGS, Abell 2744, MACSJ0647, and RXJ2129 fields. For each galaxy in the cluster fields, we corrected for lensing magnification given its sky location and spectroscopically measured redshift. We used the lensing models of \cite{2022arXiv221204381F} and \cite{2015ApJ...801...44Z} for this purpose. 

\section{Photometry Analysis} \label{sec: phot analysis}

\subsection{Stellar Population Inference} \label{sec: phot-SED}

We use the \texttt{prospector} software \citep{prospector} to infer the stellar population properties of the galaxies in our sample. \texttt{Prospector} models the observed spectral energy distribution (SED) with synthetic spectra generated using the stellar population synthesis code \texttt{FSPS} \citep{FSPS1, FSPS2}, accessed through the \texttt{python} bindings of \cite{2014zndo.....12157F}. We adopt a similar \texttt{prospector} setup to what we used in \cite{langeroodi2023} and \cite{williams+2023} for SED fitting of high-redshift galaxies with spectroscopically confirmed redshifts.

The redshift is fixed to the value measured based on emission lines. The star-formation history (SFH) is modeled nonparametrically in 5 temporal bins (see \cite{langeroodi2023} for the details of our temporal bin setup). Our stellar population free parameters include the total formed stellar mass, stellar metallicity, nebular metallicity, nebular ionization parameter, and dust attenuation. Adopting the model of \cite{2013ApJ...775L..16K}, dust attenuation is fitted with a two-component model: a diffuse dust component for the entire galaxy and a birth-cloud component for the young stars.

We use the built-in \texttt{dynesty} sampler \citep{dynesty, 2022zndo...6609296K} to explore the stellar population parameter space. \texttt{Dynesty} adopts the dynamic nested sampling method developed by \cite{2019S&C....29..891H}. We infer the best-fit parameters and their uncertainties from the weighted medians and $1\sigma$ distributions of the last 10 percentiles of \texttt{dynesty} chains, respectively. 

\subsection{Size Measurement} \label{sec: phot-size}

We measure galaxy sizes using \texttt{galight} \citep{galight}, a wrapper around the \texttt{lenstronomy} image modeling tool \citep{lenstronomy1, lenstronomy2, lenstronomy3} which automatically subtracts the sky background, identifies the bright objects in a given image cutout, and models them with user-specified light profiles. For simplicity, we model all the identified bright objects with Sersic profiles \citep{1968adga.book.....S}. \cite{2021ApJ...921...38K} and \cite{2021MNRAS.501.1028Y} have demonstrated that half-light radii measured by \texttt{galight} are consistent with those measured by \texttt{galfit} \citep{galfit}. 

More specifically, we measure the rest-UV sizes of the galaxies in our sample to target the extent of their star-forming regions. This choice was forced because, at redshifts considered in this study, the rest-UV falls in the NIRCam short wavelength filters which have much better spatial resolution and much smaller PSFs than the NIRCam long wavelength filters. Hence, targeting the rest-UV enables more accurate size measurements. We model each galaxy in the broad-band filter that covers rest-frame 2000\AA. This is mainly to ensure uniformity, and to avoid potential biases resulting from wavelength-dependent galaxy sizes. The downside is that this choice limits our analysis to galaxies at $z > 4$, below which the rest-frame 2000\AA\ falls in the NIRCam F090W and shorter-wavelength HST filters. The former does not cover a significant fraction of our sources and the latter does not provide a similar spatial resolution to NIRCam short-wavelength photometry.

Although we have constructed empirical PSFs for our entire sample (see Section \ref{sec: data-phot}), we chose to use \texttt{WebbPSF} \citep{wPSF2, wPSF3} PSFs for light profile modeling. This choice ensures that the sizes are measured consistently across our entire sample, especially given that the empirical PSFs can be under-sampled for the fields which are only covered in a few NIRCam pointings (e.g., programs 1433 and 2767). \cite{Baggen+2023} have shown that the measured Sersic half-light radii are not sensitive to the choice of PSF and that the empirical and \texttt{WebbPSF} PSFs perform similarly. 

\section{Spectral Analysis} \label{sec: spec analysis}

\subsection{Emission Line Measurement} \label{sec: spec-lines}

We use the Penalized PiXel-Fitting package \citep[\texttt{pPXF;}][]{pPXF1, pPXF2, pPXF3} to measure the emission line fluxes from the calibrated optimally-extracted 1D spectra. \texttt{pPXF} simultaneously models the continuum and emission lines by fitting the former with a stellar population and the latter with Gaussian profiles. We adopt the built-in MILES stellar library \citep{MILES1, MILES2}.

We correct for intrinsic dust attenuation as follows. If both the H$\beta$ and H$\alpha$ line fluxes are measured with high confidence (S/N $> 5$), the Balmer decrement \citep[assuming Case B recombination;][]{1989agna.book.....O} is used to correct the measured line fluxes for dust reddening adopting a routine similar to that detailed in \cite{SCALIBR}. Otherwise, we use the best-fit $A_{\rm V}$ of the diffuse dust component, as inferred by SED fitting (see Section \ref{sec: phot-SED}), to correct for dust attenuation. In both cases we adopt a \cite{2000ApJ...533..682C} dust curve. Prior to correcting for intrinsic dust attenuation, the measured line fluxes are also corrected for Galactic extinction and lensing magnification following the prescriptions described in Section \ref{sec: data-spec}.

We measure the SFR of each galaxy from its H$\alpha$ line flux, where available. We use the relation from \cite{2013seg..book..419C} to convert the measured H$\alpha$ flux to SFR. At redshifts above $z \sim 7$, where H$\alpha$ falls out of the NIRSpec prism coverage, we use the H$\beta$ flux measurements to estimate the H$\alpha$ fluxes (assuming Case B recombination) and measure the SFRs. The measured SFRs are converted to Chabrier IMF \citep{chabrier+2003} to be consistent with stellar masses inferred by SED-fitting. 

The adopted metallicity diagnostic method in this work (see Section \ref{sec: spec-metallicity}) relies on the equivalent width of H$\beta$ (EW$_{{\rm H}\beta}$) as a proxy for determining the ionization state of ISM. We measure EW$_{{\rm H}\beta}$ using the best-fit Gaussian profile to H$\beta$ and the best-fit stellar continuum, both retrieved from the \texttt{pPXF} fits. The EW is measured in the wavelength range out of which the line flux has fallen below 0.01 of its peak flux.  

\subsection{Metallicity Measurement} \label{sec: spec-metallicity}

We measure the gas-phase metallicities of the galaxies in our sample by closely following the empirical method detailed in \cite{Nakajima+2022, Nakajima+2023}. We use the strong emission line vs. gas-phase metallicity empirical calibration from \cite{Nakajima+2022}, where the sensitivity of the calibration to the ionization state of ISM is taken into account by using EW$_{{\rm H}\beta}$ as a proxy for the ionization state of ISM \cite[see Figure 6 in][for a detailed discussion of the validity of EW$_{{\rm H}\beta}$ as a proxy for ISM's ionization state]{SCALIBR}. Depending on the measured EW$_{{\rm H}\beta}$, the strong line metallicity calibration is divided into three ionization state branches: EW$_{{\rm H}\beta} < 100$\AA, $100$\AA $<$ EW$_{{\rm H}\beta} < 200$\AA, and $300$\AA $<$ EW$_{{\rm H}\beta}$. We use the EW$_{{\rm H}\beta}$ values measured in Section \ref{sec: spec-lines} to decide which branch of the calibration applies to each galaxy.

For each galaxy, we use the R23\footnote{([\ion{O}{3}]$\lambda\lambda4959,5007$+[\ion{O}{2}]$\lambda\lambda3727,29$)/H$\beta$} ratio to determine its gas-phase metallicity. Each R23 ratio corresponds to two metallicity solutions. If these solutions are not sufficiently separated (i.e., they are within $1\sigma$ uncertainty of one another), we use the lower limit of the lower metallicity solution and the upper limit of the higher metallicity solution as the $1\sigma$ uncertainty region of the measured metallicity (with nominal value defined as the mean of the two solutions). If the two metallicity solutions are sufficiently separated, we use the monotonic O32\footnote{[\ion{O}{3}]$\lambda5007$/[\ion{O}{2}]$\lambda\lambda3727,29$} ratio to differentiate between them, choosing the solution for which the expected O32 value is closer to the observed value. Objects for which R23 and O32 ratios could not be constrained with S/N $> 2$ are excluded from our analysis throughout the rest of this work. We note that in the low-metallicity regime (i.e., 12 + $\log$(O/H) $< 7.6$), metallicities measured using the calibration of \cite{Nakajima+2022} are in excellent agreement with those measured using the calibration of \cite{izotov+2019}. 

\section{Evolution of Galaxy Sizes} \label{sec: size evol}

\subsection{Mass-Size Relation} \label{sec: size-MS}

\begin{figure}
    \centering
    \includegraphics[width=8.5cm]{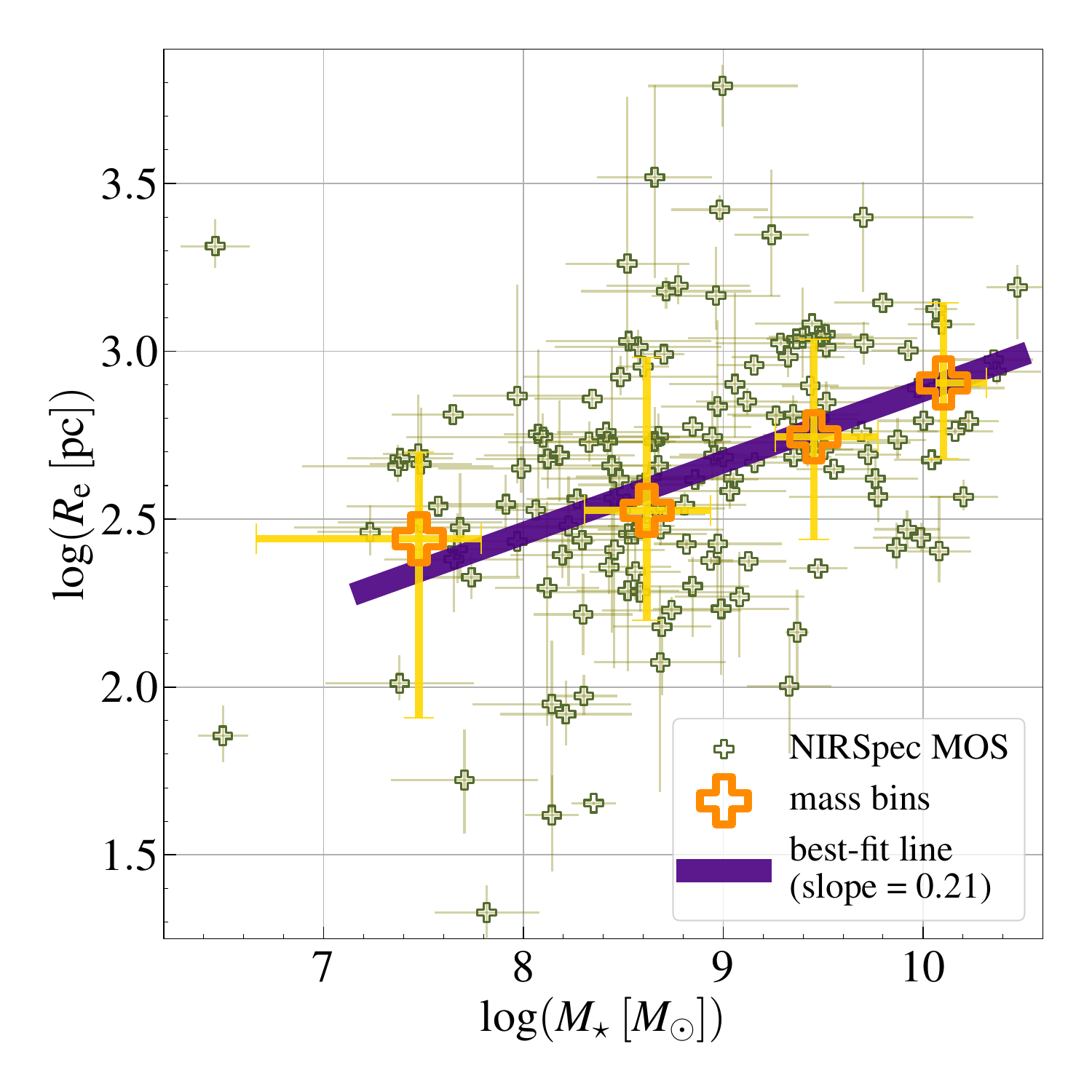}
    \caption{Mass-size relation at $z = 4-10$. The small green data points show the measured rest-UV size and stellar mass for the galaxies in our sample. The large orange data points show the weighted median and $1\sigma$ distribution of the data in four stellar mass bins. The dark purple line shows the best-fit power law mass-size relation with a slope of $0.21 \pm 0.04$.}
    \label{fig: mass-size}
\end{figure}

Figure \ref{fig: mass-size} shows the correlation between rest-UV size and stellar mass for the $4 < z < 10$ galaxies in our sample. The large orange data points show the weighted median and $1\sigma$ distribution of the data in four stellar mass bins. We fit the mass-size relation through least-squared fitting; the dark purple line shows the best-fit power law of the form
\begin{align}
    & \log(R_{\rm e} {\rm [pc]}) = \notag \\ 
    & (0.21 \pm 0.04) \log(M_{\star}/M_{\odot}) + (0.75 \pm 0.38) \;.
\end{align}
Interestingly, the highly magnified $z_{\rm spec} = 9.51$ galaxy found behind the lensing cluster RXJ2129 is the smallest and most compact entry in our sample, with $R_{\rm e} \sim 20 {\rm pc}$ at $\log(M_{\star}/M_{\odot}) \sim 7.8$ \citep{williams+2023}.

\begin{figure}
    \centering
    \includegraphics[width=8.5cm]{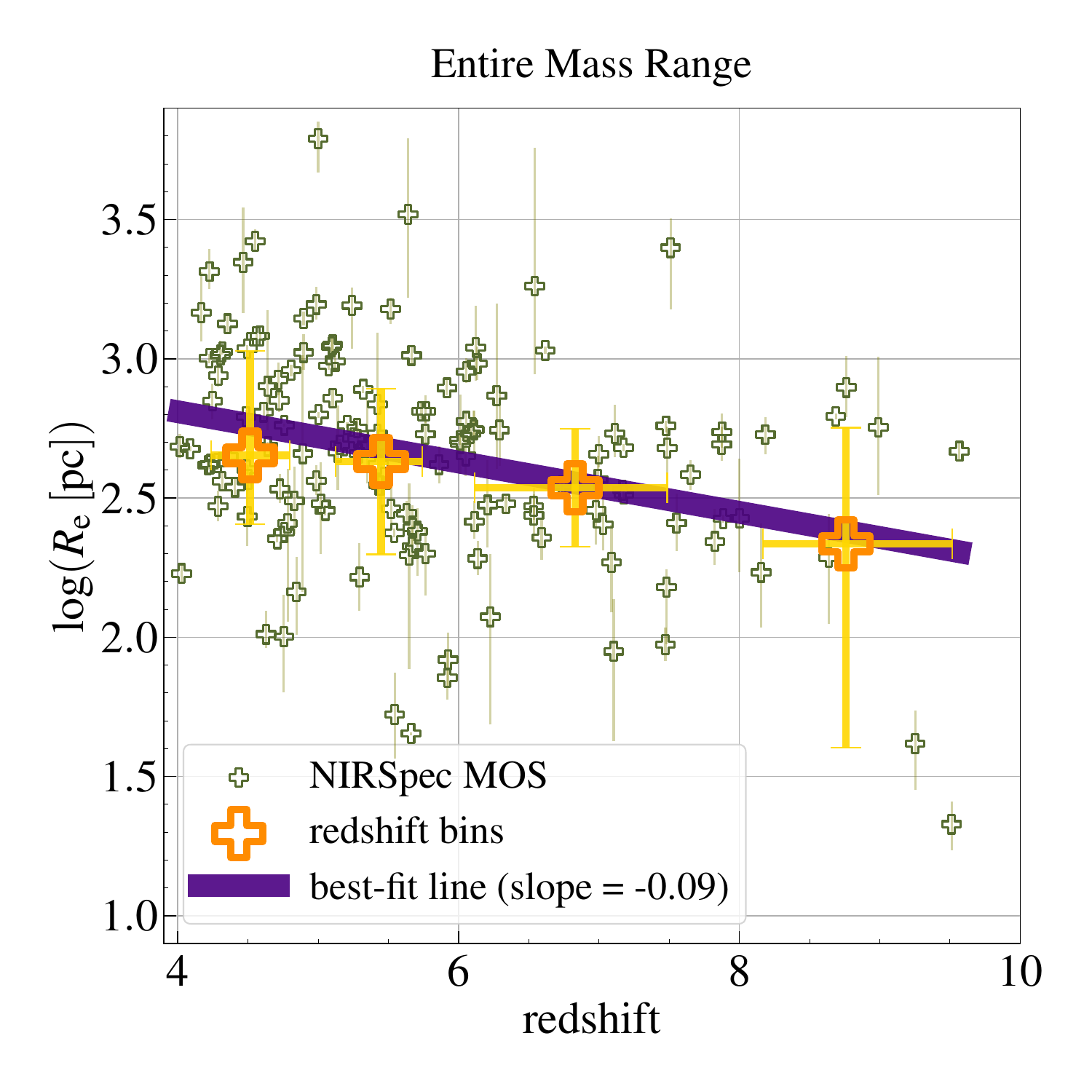}
    \caption{Redshift evolution of the mass-size relation. The small green data points show the measured rest-UV size vs. spectroscopic redshift for the galaxies in our sample. The large orange data points show the weighted median and $1\sigma$ distribution of the data in four redshift bins. There is a clear redshift trend, where galaxies get increasingly more compact at higher redshifts. The dark purple line shows the best-fit line with a slope of $-0.090 \pm 0.02$ and an intercept of $3.17 \pm 0.14$.}
    \label{fig: size-redshift all}
\end{figure}

\begin{figure*}
    \centering
    \includegraphics[width=\linewidth]{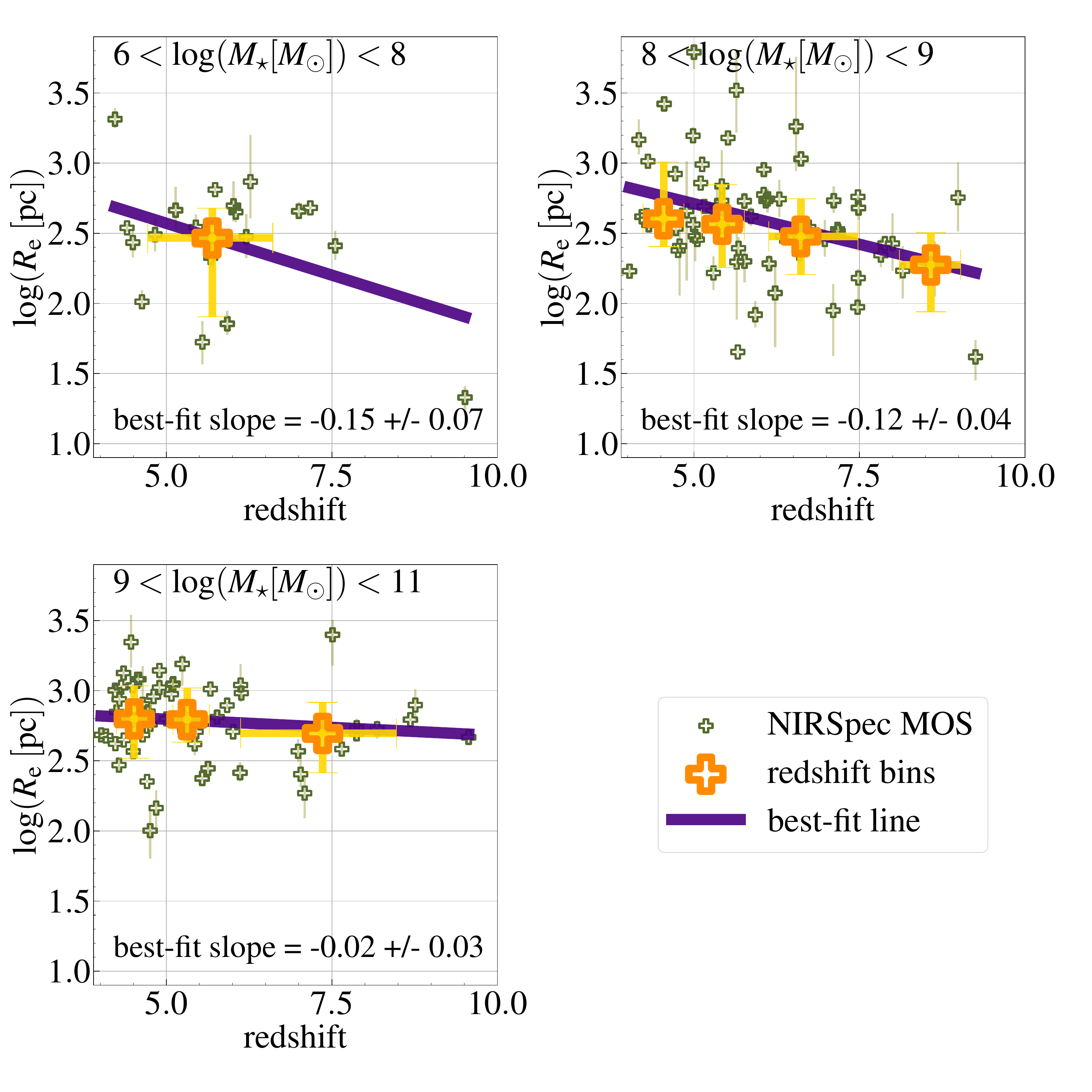}
    \caption{Redshift evolution of the mass-size relation in bins of stellar mass. The small green data points in each panel show the measured rest-UV size vs. spectroscopic redshift for galaxies in a limited stellar mass bin, indicated at the top of the panel. The large orange data points show the weighted median and $1\sigma$ distribution of the data in redshift bins. The dark purple lines show the best-fit size-redshift linear relations, with slopes noted at the bottom of the corresponding panel. The lowest-mass galaxies ($M_{\star} < 10^{8} M_{\odot}$; upper-left panel) show the strongest size-redshift evolution, while the highest-mass galaxies ($10^{9} M_{\odot} < M_{\star}$; lower-left panel) show no noticeable redshift evolution.}
    \label{fig: size-redshift bins}
\end{figure*}

\begin{figure*}
    \centering
    \includegraphics[width=0.9\linewidth]{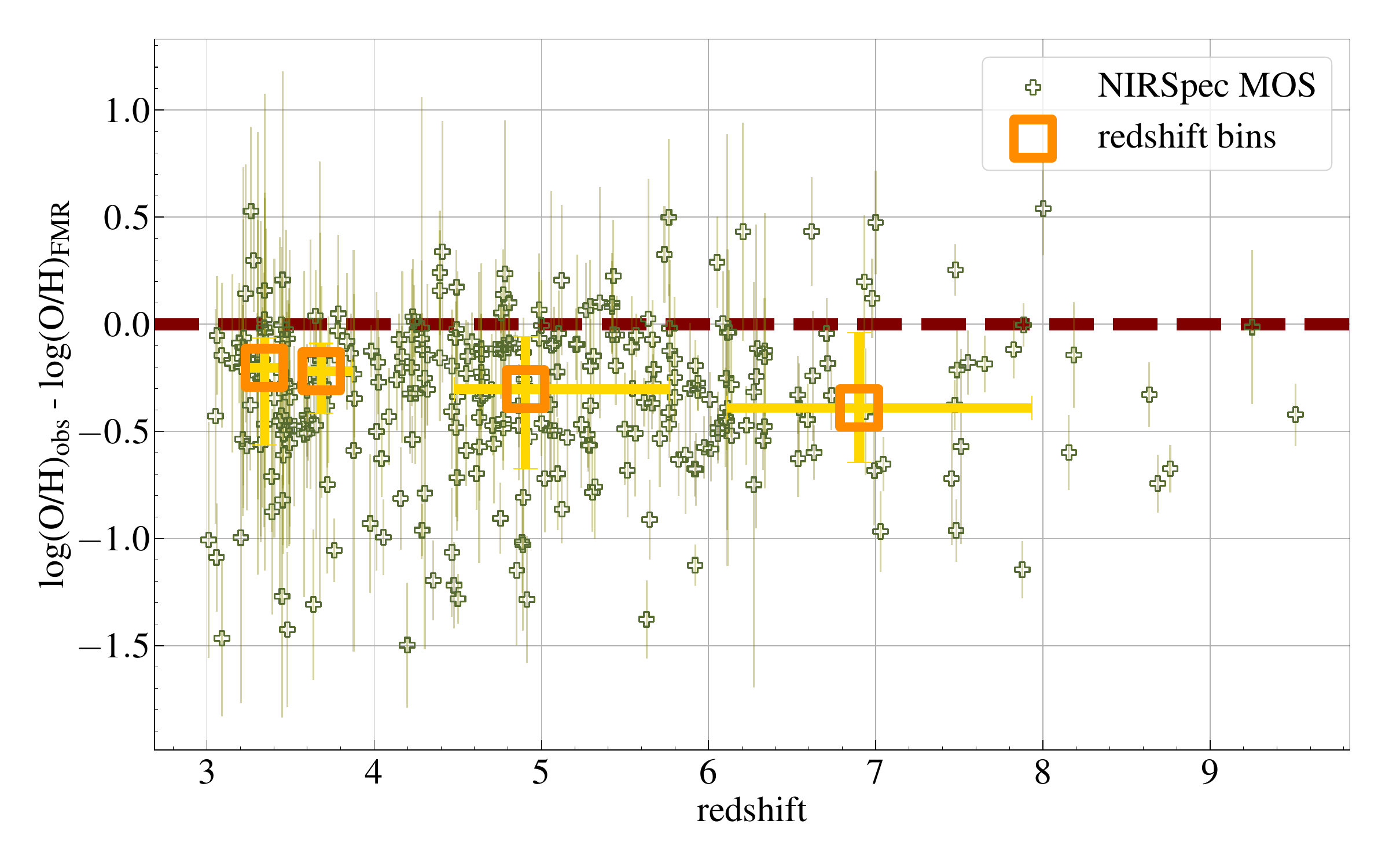}
    \caption{Redshift evolution of the fundamental metallicity relation. The small green data points show the offsets of the measured metallicities for the galaxies in our sample from those expected for their stellar masses and SFRs based on the local universe calibration of the FMR. The large orange data points show the weighted medians and $1\sigma$ distributions of the data in bins of redshift. The dashed red line indicates zero offset from the local universe calibration of the FMR. There is a mild redshift evolution in the FMR.}
    \label{fig: FMR}
\end{figure*}

\begin{figure}[b]
    \centering
    \includegraphics[width=1\linewidth]{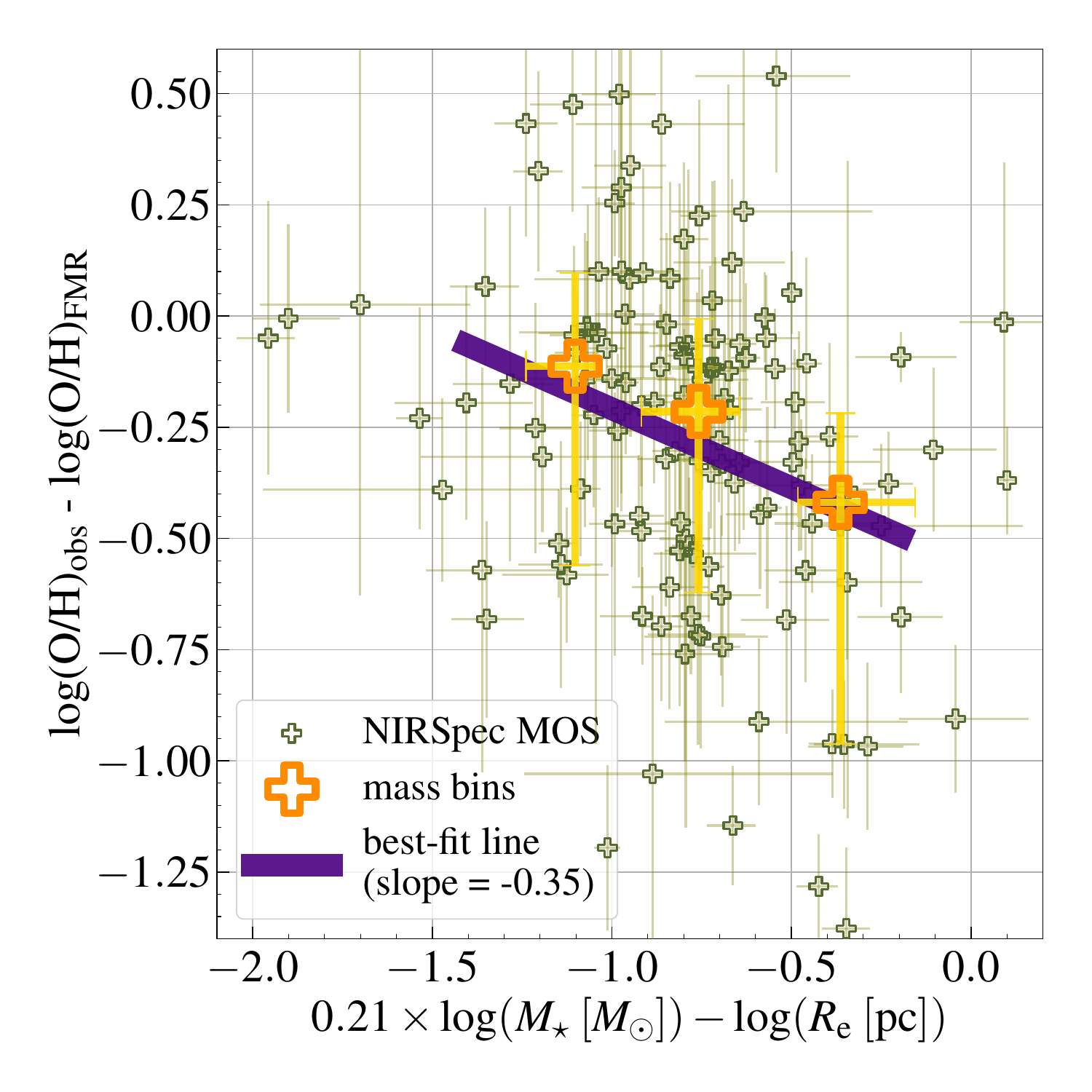}
    \caption{Compact galaxies are more metal-deficient. The small green data points show the offset from the local universe calibration of the FMR vs. compactness for the galaxies in our sample. Compactness is defined as $0.21\times\log(M_{\star}) - \log(R_{\rm e})$, which essentially captures the deviation from the average mass-size relation (see Section \ref{sec: FMR-compactness}). In simple words: along the y-axis, galaxies become more metal-rich than expected; while along the x-axis, galaxies become more compact. The large orange data points show the weighted medians and $1\sigma$ distributions of the data in bins of compactness. The purple line shows the best-fit relation of the form given by Equation \ref{eq: offset vs compactness}.}
    \label{fig: FMR-compactness}
\end{figure}

\subsection{Redshift Evolution of the Mass-Size Relation} \label{sec: size-SR}

Next, we investigate the redshift evolution of the mass-size relation. 
Figure \ref{fig: size-redshift all} shows the measured size of the galaxies in our sample plotted against their spectroscopic redshifts (small green data points). The large orange data points show the weighted median and $1\sigma$ distribution of the data in four redshift bins. There is a clear redshift trend, where galaxies become progressively more compact with increasing redshift. We fit for this trend through least-squared fitting; the dark purple line shows the best-fit line of the form 
\begin{equation}
\log(R_{\rm e} {\rm [pc]}) = (-0.090 \pm 0.02) z + (3.17 \pm 0.14) \;.   
\end{equation}

The inferred general size-redshift anti-correlation might be misleading, as it is likely affected by two mass-related systematics. First; the number density of high-mass galaxies is expected to decline with redshift. Since the mass-size relation implies that these are the largest galaxies, a decline in their number density can exaggerate the inferred size-redshift anti-correlation. Second; due to detection limits, higher-redshift samples are expected to be biased towards the brightest and most massive galaxies. This can artificially boost the average size of high-redshift samples, leading to a flattened size-redshift trend. The described systematics work in opposite directions, as one steepens the size-redshift anti-correlation and one flattens it. However, their combined effect cannot be neglected.

In order to mitigate these systematics, we consider the size-redshift anti-correlation in bins of stellar mass. This is shown in Figure \ref{fig: size-redshift bins}, where the upper-left, upper-right, and lower-left panels respectively correspond to the $\log(M_{\star}/M_{\odot}) < 8$, $8 < \log(M_{\star}/M_{\odot}) < 9$, and $9 < \log(M_{\star}/M_{\odot})$ galaxies. The large orange data points in each panel show the median and $1\sigma$ distribution of the data in redshift bins. The dark purple line in each panel shows the best-fit least-squared size-redshift relation in the corresponding stellar mass bin; the best-fit slopes and their uncertainties are noted at the bottom of each panel. 

Figure \ref{fig: size-redshift bins} presents a consistent picture: galaxies become progressively more compact with increasing redshift. The strength of this redshift evolution depends on the stellar mass. The lowest mass galaxies ($M_{\star} < 10^{8} M_{\odot}$) exhibit the strongest redshift evolution, while the sizes of highest mass galaxies ($10^{9} M_{\odot} < M_{\star}$) do not appear to noticeably evolve with redshift. 

\section{Fundamental Metallicity Relation} \label{sec: FMR}

In this Section, we investigate the redshift evolution of the FMR (Section \ref{sec: FMR-FMR}) and the role of UV-compactness in driving this redshift evolution (Section \ref{sec: FMR-compactness}). For each galaxy in our sample, we use the local universe calibration of the FMR from \cite{AM13} to derive the expected gas-phase metallicity given its stellar mass (measured through SED fitting; Section \ref{sec: phot-SED}) and SFR (measured from H$\alpha$ or H$\beta$; Section \ref{sec: spec-lines}). We adopt a similar form to that of \cite{Nakajima+2023}, where the calibration of \cite{AM13} is converted from the Kroupa IMF to the Chabrier IMF
\begin{align}
    & 12 + \log({\rm O/H}) = 0.43 \times \mu_{0.66} + 4.58\;; \\ 
    & {\rm where}\;\; \mu_{\alpha} = \log M_{\star} - \alpha \log {\rm SFR}\;.
\label{eq: FMR}
\end{align}

\vspace{7.5pt}
\subsection{Redshift Evolution of the FMR} \label{sec: FMR-FMR}

Figure \ref{fig: FMR} shows a mild redshift evolution in the FMR: the offset between the inferred gas-phase metallicities and those expected based on the local universe calibration of the FMR (i.e., $\log({\rm O/H})_{\rm obs} - \log({\rm O/H})_{\rm FMR}$) consistently increase with increasing redshift. We note that the measured offsets strongly depend on the adopted FMR calibration. For instance, \cite{Nakajima+2023} showed that adopting the calibration from \cite{curti+2020} results in more significant offsets across the entire redshift range. Nonetheless, as shown in both \cite{Nakajima+2023} and \cite{curti+2023}, the general trend of a gradually increasing metallicity offset with increasing redshift persists. 

Moreover, we note that the inferred offsets are sensitive to the adopted empirical strong-line metallicity calibration. For instance, using the calibration from \cite{sanders+2023} results in generally higher offsets (in the same direction as Figure \ref{fig: FMR}) as well as larger scatter in the offsets. Adopting the calibration from \cite{izotov+2019} for $12 + \log({\rm O/H}) < 7.6$ galaxies (the metallicity range where this calibration is valid) results in offsets that are similar to what is shown in Figure \ref{fig: FMR}, but a with much lower scatter. Nonetheless, in both scenarios, the redshift evolution persists. 

\subsection{FMR Offset vs. Compactness} \label{sec: FMR-compactness}

Figure \ref{fig: FMR-compactness} shows the FMR offset of the galaxies in our sample plotted against their ``compactness''. We define compactness as 
\begin{equation}
    {\rm compactness} = \kappa \log(M_{\star}\;[M_{\odot}]) - \log(R_{\rm e}\;[{\rm pc}]),
\label{eq: compactness}
\end{equation}
where $\kappa$ is the slope of the best-fit mass-size power law relation (see Figure \ref{fig: mass-size} and Section \ref{sec: size-MS}). Compactness essentially captures the offset from the average mass-size relation, i.e., measuring if the galaxy is smaller than expected given its stellar mass. We set $\kappa = 0.21$, based on the best-fit mass-size relation derived in Section \ref{sec: size-MS}.

Figure \ref{fig: FMR-compactness} shows that the FMR offset and compactness are correlated: the most compact galaxies are also the most offset from the local universe calibration of the FMR. Through least-squared fitting, we fit this correlation with a linear relation of the form 
\begin{align}
    & \log({\rm O/H})_{\rm obs} - \log({\rm O/H})_{\rm FMR} = \nonumber \\ 
    & (-0.35 \pm 0.11) \times {\rm compactness} -0.56 \pm 0.09 \;,
\label{eq: offset vs compactness}
\end{align}
where compactness is defined as Equation \ref{eq: compactness}, with $\kappa = 0.21$. This best-fit relation is shown as the dark purple line in Figure \ref{fig: FMR-compactness}. The correlation indicates that as galaxies get more compact at a fixed stellar mass and SFR, they become more metal-poor. We discuss the implications of this finding in Section \ref{sec: discussion}. 

\section{Discussion} \label{sec: discussion}

We interpret the observed metal-deficiency of compact galaxies as a result of bursty star formation. Since the rest-UV emission traces star-formation, by definition, these young UV-compact galaxies display elevated SFR surface densities. This suggests that they are observed at the bursty phase of their star formation histories. While ISM metals and dust are produced rapidly following an star formation event \citep{langeroodi2023, DUSTSFR}, the overall metal-deficiency of these compact galaxies strongly suggests that their burst episodes are shorter lived than their enrichment timescales. This puts strong constraints on the duty cycles of bursty star formation and metal mixing timescales in early-universe galaxies. We expand on this idea in the following. 

The FMR can be interpreted as the consequence of the role of gas infall in simultaneously enhancing star formation while lowering gas-phase metallicities \citep{2010MNRAS.408.2115M, 2019A&ARv..27....3M}. The accreted low-metallicity gas lowers gas-phase metallicity by diluting the ISM but enhances SFR by providing fuel for star formation. As such, an anti-correlation between the SFR and gas-phase metallicity is expected at a fixed stellar mass; this is known as the FMR. The shape and redshift evolution of the FMR can place fundamental constraints on the interplay of gas infall, enrichment, and gas ejection as well as their corresponding timescales. 

More precisely, the FMR can be interpreted as an equilibrium state between gas infall, enrichment, and gas ejection. In this context, the UV-compact bursty galaxies observed in the early-universe have not yet settled into equilibrium following their burst episodes. In other words, these galaxies are caught at a phase where their SFR is enhanced and their gas-phase metallicity is lowered following an intense gas infall and burst episode, but before the chemical yield from the newly formed massive stars is added to the ISM. In other words, they are chemically out of equilibrium. 

This picture is fully consistent with the predicted burst cycles in high-redshift and/or low-mass galaxies based on zoom-in simulations \citep[see e.g.,][for an overview]{2023MNRAS.525.2241H}, where feedback from newly formed massive stars competes with the gravitational potential to disperse cold gas and suppress further star formation. Supernova feedback as well as faster processes such as stellar winds and radiation from massive stars contribute to this self-regulation. Chemical enrichment, at best, operates on the same timescale as supernova feedback (i.e., making the instantaneous recycling approximation). Our observations suggest that bursts are suppressed on faster timescales, driven by radiation and stellar winds. This is in line with recent observational evidence concluding that pre-supernova feedback mechanisms are primarily responsible for dispersing giant molecular clouds on spatial scales relevant to most of our compact high-redshift galaxies \citep[up to a few hundred pc;][]{2019Natur.569..519K, 2022MNRAS.509..272C}.

\section{Conclusion} \label{sec: conclusion}

In this work, we compiled a sample of 427 galaxies at $3 < z_{\rm spec} < 10$, all covered by both the NIRSpec MSA prism spectroscopy and NIRCam short-wavelength photometry. The spectra consist of our uniform reduction of publicly available NIRSpec spectroscopy for 334 galaxies as well as 93 galaxies from the first data release of the JADES program. The emission lines were measured using \texttt{pPXF}. We uniformly measured the photometry for the entire sample of 427 galaxies after PSF-matching and aperture correction. We measured the stellar population, nebular, and morphological properties of these galaxies using \texttt{prospector} SED-fitting, strong-line metallicity estimation, and \texttt{galight} light profile fitting. Our main findings are as follows.

We find that a mass-size power law relation is already present for the $4 < z < 10$ galaxies in our sample over more than 3 decades in stellar mass: the most massive galaxies appear to be the most extended. We also find a general decline in the size of average galaxies with redshift: $z \sim 10$ galaxies are on average $\sim 4$ times smaller than $z \sim 4$ galaxies. We divide our sample into three stellar mass bins, confirming that the size-redshift anti-correlation persists regardless of stellar mass. In other words, we find that at a fixed stellar mass higher redshift galaxies are on average less extended. The strength of the size-redshift anti-correlation depends on stellar mass: sizes of low-mass galaxies exhibit strong redshift dependence, whereas those of massive galaxies exhibit negligible redshift evolution.

We find a mild redshift evolution of the FMR. The inferred gas-phase metallicities of $z > 3$ galaxies are offset from the local-universe FMR expectations for their stellar mass and SFR. In other words, early-universe galaxies appear more metal-poor than expected. We also find that this metal-deficiency consistently grows with redshift from $z \sim 3$ to $z \sim 10$. While the FMR has been quite successful at explaining the redshift evolution of the normalization of mass-metallicity relation out to cosmic noon, our findings emphasize that the FMR fails to fully capture the evolution beyond these redshifts. 

We find that metal-deficiency is correlated with compactness: the most compact galaxies are also the most metal deficient by being the most offset from the local-universe calibration of the FMR. We interpret this correlation as a consequence of bursty star formation in these early-universe galaxies. The compact galaxies exhibit elevated SFR surface densities, indicating that they are observed during their burst episodes induced by gas-accretion events. While the accretion of metal-poor gas has reduced their gas-phase metallicity by diluting the ISM, these compact galaxies are observed prior to the chemical yield release by newly formed massive stars. In other words, they are chemically out of equilibrium from the equilibrium state that is observed in the local Universe as the FMR. 

\section{Acknowledgments}

D.L. greatly appreciates helpful discussions with Mirko Curti, Kimihiko Nakajima, and Ryan Sanders which influenced some of the interpretations and conclusions of this work. We are thankful for the technical comments by Pablo Pérez-González on PSF-matching and flux uncertainty measurements. We are indebted to Gabe Brammer for making his JWST and HST photometry reductions publicly available. We acknowledge the clear description of the post-pipeline NIRCam photometry analysis by the CEERS team and Yuichi Harikane in their respective papers, without which those steps of this work would have become much more challenging. This work was supported by research grants (VIL16599, VIL54489) from VILLUM FONDEN.

\vspace{10pt}
Software:
dynesty \citep{dynesty},
EAZY \citep{eazy}, 
FSPS \citep{FSPS1, FSPS2},
galight \citep{galight}, 
Grizli \citep{grizli},
msaexp \citep{msaexp},
pPXF \citep{pPXF1, pPXF2, pPXF3}, 
PSFex \citep{psfex},
prospector \citep{prospector},
sedpy \citep{sedpy}, 
Source Extractor \citep{sex},
WebbPSF \citep{wPSF2, wPSF3}.

\bibliography{main}
\bibliographystyle{aasjournal}

\end{document}